\documentstyle[prl,aps,twocolumn]{revtex}
\begin{document}
\voffset 0.5in
\draft
\wideabs{
\title{Spin currents and spin dynamics in time-dependent density-functional 
theory}
\author{K. Capelle}
\address{Departamento de Qu\'{\i}mica e F\'{\i}sica Molecular,
Instituto de Qu\'{\i}mica de S\~ao Carlos,
Universidade de S\~ao Paulo,\\
Caixa Postal 780, S\~ao Carlos, 13560-970 SP,
Brazil}
\author{G. Vignale}
\address{
Department of Physics and Astronomy, University of Missouri-Columbia,
Columbia, Missouri 65211, USA}
\author{B. L. Gy\"orffy}
\address{H.H. Wills Physics Laboratory, University of Bristol, Bristol BS8 1TL,
United Kingdom}
\date{\today}
\maketitle
\begin{abstract}
We derive and analyse the equation of motion for the spin degrees of freedom
within time-dependent spin-density-functional theory (TD-SDFT). 
Results are (i) a prescription for obtaining many-body corrections to the 
single-particle spin currents from the Kohn-Sham equation of TD-SDFT,
(ii) the existence of an exchange-correlation (xc) torque within TD-SDFT,
(iii) a prescription for calculating, from TD-SDFT, the torque exerted by 
spin currents on the spin magnetization,
(iv) a novel exact constraint on approximate xc functionals, and
(v) the discovery of serious deficiencies of popular approximations to TD-SDFT 
when applied to spin dynamics.
\end{abstract}

}
\newcommand{\be}{\begin{equation}}
\newcommand{\ee}{\end{equation}}
\newcommand{\bea}{\begin{eqnarray}}
\newcommand{\eea}{\end{eqnarray}}
\newcommand{\bi}{\bibitem}

\newcommand{\ep}{\epsilon}
\newcommand{\s}{\sigma}
\newcommand{\p}{{\bf \pi}}
\newcommand{\r}{({\bf r})}
\newcommand{\x}{({\bf x})}
\newcommand{\xt}{({\bf x},t)}
\newcommand{\rt}{({\bf r},t)}
\newcommand{\rit}{({\bf r}_i,t)}
\newcommand{\rp}{({\bf r'})}
\newcommand{\rrp}{({\bf r},{\bf r'})}
\newcommand{\rpt}{({\bf r}',t)}
\newcommand{\rpp}{({\bf r''})}

\newcommand{\ua}{\uparrow}
\newcommand{\da}{\downarrow}
\newcommand{\la}{\langle}
\newcommand{\ra}{\rangle}
\newcommand{\dg}{\dagger}

The dynamics of the spin degrees of freedom is responsible for 
such diverse phenomena as spin wave excitations, Bloch wall motion, 
spin-polarized currents, and spin injection and spin filtering; 
concepts and phenomena which are important, e.g., in the growing field 
of {\it spintronics} \cite{spindyn}.
The calculation of spin dynamics within density-functional theory (DFT)
has consequently received much 
attention \cite{calc}.
The most popular DFT method for a first-principles treatment of the
spin degrees of freedom is spin-density-functional theory 
(SDFT, see Refs.~\onlinecite{dftbook} and \onlinecite{joubert} for reviews.)
SDFT has led to versatile and powerful schemes for the calculation 
of, e.g., total energies, spin densities, and spin-resolved single-particle 
bandstructures, but its traditional (i.e., ground state)
formulation is applicable only to static situations. 
This situation has changed with the advent of time-dependent DFT (TD-DFT)
\cite{tddft}, which has brought dynamical phenomena 
within reach of DFT.

In the present paper we first derive the equation of motion for the 
spin magnetization from TD-SDFT, including exchange-correlation (xc) 
effects. We then show how this equation can be used to obtain information
on the many-body spin current. Although one might
think that a calculation of spin {\it currents} would require the
more complex formalism of time-dependent {\it current}-density-functional 
theory (CDFT), we find that this is not entirely true: as a consequence of 
the continuity equation TD-SDFT suffices to calculate the 
spin currents in several cases of great of practical interest.

Numerical applications of TD-SDFT, as of any other DFT, require
knowledge of the xc potentials, which contain all
many-body effects beyond the Hartree approximation. In traditional DFT many
exact properties of these xc potentials are known, and greatly aid the
construction of good approximations \cite{dftbook,joubert}, but the same is 
not true in the time-dependent case, where properties 
of the xc potentials are just beginning to be explored \cite{conds}.
As a byproduct, our analysis reveals a previously unknown exact property of 
the xc potentials of TD-SDFT, which strongly constrains suitable approximations.

In TD-SDFT the fundamental variables are the 
time-dependent particle density,
\be
n\rt = \sum_i^N\left\langle \Psi\left|
\delta({\bf r}-{\bf r}_i)
\right|\Psi\right\rangle
\label{ndef}
\ee
and the time-dependent magnetization (or spin) density 
\be
{\bf m}\rt = \mu_0 \sum_i^N \left\langle \Psi\left|
\hat{\bbox{\s}}_i \delta({\bf r}-{\bf r}_i)
\right|\Psi\right\rangle,
\label{mdef}
\ee
where $\Psi=\Psi({\bf r}_1,..,{\bf r}_N,t)$ is the many-body
wave function (spin coordinates are suppressed for 
brevity), $\mu_0 = q\hbar/(2mc)$ is the Bohr magneton, $\hat{\bbox{\s}}_i$ is 
the vector of Pauli matrices, and $N$ the particle number.
Here and below all operators are taken to be independent of time,
i.e., the time dependence of expectation values 
results exclusively from that of the wave function.
In TD-SDFT these expectation values are not calculated with the 
many-body wave function $\Psi$, but from the solutions of a noninteracting 
Hamiltonian, containing suitably choosen effective electrostatic
and magnetic fields $v_s\rt$ and ${\bf B}_s\rt$, 
\bea
\hat{H}^{KS}=
\sum_i^N\left[-\frac{\hbar^2\nabla_i^2}{2m} + v_s\rit -
\mu_0 \hat{\bbox{\s}}_i\cdot{\bf B}_s\rit
\right].
\label{tdsdftks}
\eea

Eq.~(\ref{tdsdftks}) defines the Kohn-Sham (KS) Hamiltonian of TD-SDFT.
The solution of $[i\hbar \partial/(\partial t) -\hat{H}^{KS}]\Phi=0$, i.e., 
the Slater determinant $\Phi({\bf r}_1,..,{\bf r}_N,t)$, 
reproduces the correct particle 
and spin densities as functions of both time and position. 
By contrast, the current
\be
\hat{\bf j}\r =
\frac{\hbar}{2mi}\sum_i^N
\nabla_i\delta({\bf r}-{\bf r}_i)+\delta({\bf r}-{\bf r}_i)\nabla_i,
\label{jpara}
\ee
is not among the basic variables of TD-SDFT, 
so that ${\bf j}\rt=\langle \Psi|\hat{\bf j}\r|\Psi\rangle
\neq \langle \Phi|\hat{\bf j}\r|\Phi\rangle ={\bf j}^{KS}\rt$.

The equations of motion for the fundamental density variables 
describing the electronic degrees of freedom, $n\rt$ and ${\bf m}\rt$,
can be calculated directly from the commutator with the Kohn-Sham 
Hamiltonian. For the time evolution of ${\bf m}\rt$ one finds
\be
\frac{d {\bf m}\rt}{d t} 
+\hat{\nabla}\cdot {\bf J}^{KS}\rt
=\frac{q}{2mc}{\bf m}\rt  \times {\bf B}_s\rt.
\label{tdsdftlifshitz}
\ee
The second term on the left-hand side, which stems from the commutator with 
the kinetic energy, describes spin currents arising because 
the electrons carry their spin with them as they move around. Here
\be
{\bf J}^{KS}\rt:= \mu_0 \sum_i^N \langle \Phi|
\hat{\bbox{\s}}_i\otimes \hat{\bf j}_i\r 
|\Phi \rangle
\label{Jpdef}
\ee
is the KS spin-current tensor, defined via the
tensor product of the spin vector $\hat{\bbox{\s}}_i$ and the 
orbital current $\hat{\bf j}_i\r$ of particle $i$ \cite{footnote1}.
(We use a capital ${\bf J}$ to denote spin-current tensors, and a lower case
${\bf j}$ for the corresponding orbital currents.)
Eq.~(\ref{tdsdftlifshitz}) is a formally exact TD-SDFT representation of the 
time evolution of the spin degrees of freedom. Many-body effects enter
Eq.~(\ref{tdsdftlifshitz}) via the effective magnetic field
${\bf B}_s$, which is defined as 
${\bf B}_s\rt = {\bf B}\rt + {\bf B}_{xc}\rt$,
where ${\bf B}\rt$ is the external magnetic field, and ${\bf B}_{xc}\rt$
the exchange-correlation magnetic field.

Of course, the equation of motion for ${\bf m}\rt$ can also be derived 
from the many-body Hamiltonian. Since in the absence of relativistic effects 
the spin magnetization commutes with the particle-particle interaction, the
result takes the same form as in the Kohn-Sham system,
\be
\frac{d {\bf m}\rt}{d t}
+\hat{\nabla}\cdot {\bf J}\rt 
=\frac{q}{2mc}{\bf m}\rt  \times {\bf B}\rt,
\label{manybody}
\ee
up to the replacement of the effective by the external magnetic field,
and the Kohn-Sham current ${\bf J}^{KS}$ by the many-body current 
${\bf J}$.
Eq.~(\ref{manybody}) is simply the continuity equation for the 
spin magnetization.
Comparing Eqs.~(\ref{tdsdftlifshitz}) and (\ref{manybody}) one finds that
\be
\frac{q}{2mc}{\bf m}\rt  \times {\bf B}_{xc}\rt = 
\hat{\nabla}\cdot \left[ {\bf J}^{KS}\rt - {\bf J}\rt \right]
\label{ltt}
\ee
for all times $t$ and at every point ${\bf r}$. 
This equation is the central result of the present analysis.
We now proceed to explore some of the consequences, practical and
fundamental, of this result.

(i) The right-hand side of Eq.~(\ref{ltt}) contains the difference between 
the many-body spin current and its
Kohn-Sham counterpart. This difference is in the following denoted
${\bf J}_{xc}$, since it is the exchange-corrrelation contribution to the
full current. {\it The longitudinal part of ${\bf J}_{xc}$, denoted 
${\bf J}^L_{xc}$, 
can be calculated directly from TD-SDFT}, by integrating Eq.~(\ref{ltt}).
It is interesting to note that this many-body correction to the Kohn-Sham
current follows from SDFT alone, without having to use CDFT. 

${\bf J}^L_{xc}$ on its own is valuable information, 
but a full determination of the many-body current requires knowledge
of the transverse correction, too, and this can, at present, only be calculated 
within CDFT. For a large class of systems, however,
${\bf J}^L_{xc}$ is the most relevant many-body correction.
This includes all quasi one-dimensional systems (e.g., conducting
polymers), in which only the longitudinal component of the spin current
is of interest, but also three-dimensional systems in which the current
flows entirely in one direction, e.g., perpendicular to an interface.
A particularly important example is provided by magneto-resistive devices
in which the spin current flow is perpendicular to the planes of a magnetic
multilayer: such a current is purely longitudinal and can therefore be
computed from Eq.~(\ref{ltt}).

(ii) The left-hand side of Eq.~(\ref{ltt}) is the torque locally exerted by the
xc magnetic field on the spin configuration.
The existence of this torque explains how TD-SDFT achieves a formally correct 
description of spin
dynamics via Eq.~(\ref{tdsdftlifshitz}) although the currents the spins are 
coupled to in that equation are the KS currents, and not the physical ones: 
{\it the exchange-correlation torque (\ref{ltt}) accounts for the difference 
between the many-body and the KS currents} in Eq.~(\ref{tdsdftlifshitz}). 
This torque arises from the component of ${\bf B}_{xc}$ perpendicular to 
${\bf m}\rt$, 
\be
|{\bf B}^\perp_{xc}\rt| = \frac{2mc}{|q|} 
\frac{|\hat{\nabla}\cdot\left[{\bf J}^{KS}\rt - {\bf J}\rt \right]|}
{|{\bf m}\rt|}.
\label{bperp}
\ee

(iii) Eq.~(\ref{manybody}) shows that for a spontaneously magnetic system 
(with ${\bf B}\rt=0$) the quantity $-\hat{\nabla}\cdot {\bf J}\rt$ can be 
interpreted as
the {\it torque exerted by the current on the spin magnetization}. 
There has recently been much interest in such torques in connection with 
experiments on current-driven magnetization reversals \cite{magrev}.
Our Eq.~(\ref{ltt}) provides a way to calculate these torques from the
SDFT quantities ${\bf B}_{xc}\rt$, ${\bf m}\rt$, and ${\bf J}^{KS}\rt$.

(iv) By integrating Eq.~(\ref{ltt}) over all space one obtains
\be
\int d^3r\, {\bf m}\rt  \times {\bf B}_{xc}\rt \equiv 0,
\label{ztt}
\ee
since the right-hand side vanishes by virtue of Gauss' theorem.
Eq.~(\ref{ztt}) has a simple physical interpretation:
{\em the self-consistent xc magnetic field cannot exert a
net torque on the system as a whole.} 
We refer to Eq.~(\ref{ztt}) in the following as the
zero-torque theorem (ZTT). This terminology puts it on the same footing as
the zero-force theorem \cite{tddft,conds} which states that the
electrostatic xc potential does not exert a net force on the system.

By integrating Eq.~(\ref{ltt}) over the volume of a finite system one finds
instead of Eq.~(\ref{ztt})
\be
\frac{q}{2mc}\int_\Omega d^3r\, {\bf m}\rt  \times {\bf B}_{xc}\rt 
=\int_S d{\bf S}\cdot {\bf J}_{xc}\rt,
\label{finiteztt}
\ee
where $\Omega$ is the volume of the system bounded by the surface ${\bf S}$.
{\it For a finite system the integral over ${\bf m}\times {\bf B}_{xc}$ thus
measures the net flux of xc spin current into and out of the system.}

(v) Finally, through Eq.~(\ref{ztt}) Eq.~(\ref{ltt}) provides a
stringent constraint on approximations for ${\bf B}_{xc}$. Any such
approximation which does not satisfy Eq.~(\ref{ztt}) is not consistent
with the microscopic spin dynamics and the continuity equation for
the spin magnetization. 
In the popular adiabatic local-density approximation (ALDA), for example, one
takes as ones approximation the xc fields of static SDFT within the usual LDA,
but evaluates these at the time dependent densities \cite{tddft}.
Since in LDA ${\bf B}_{xc}$ and ${\bf m}$ are by construction always parallel
one has, in the ALDA, ${\bf m}\rt  \times {\bf B}_{xc}\rt \equiv 0$ for all
times $t$ and at any point ${\bf r}$, and Eq.~(\ref{ztt}) is trivially 
satisfied.
The fact that the ALDA thus (fortuitiously) satisfies the ZTT
may explain the relative success LDA-based calculations have had in 
previous calculations of spin dynamics and other dynamical phenomena 
in TD-SDFT.

On the other hand, from Eq.~(\ref{ltt}) one finds that within
the ALDA ${\bf J}_{xc}^L\rt \equiv 0$.
Since one does not expect that many-body corrections to the longitudinal 
spin current vanish generally, their absence in the ALDA must be considered 
a defect of this approximation.
Furthermore, from the equation of motion (\ref{tdsdftlifshitz}) 
it follows that in the ALDA the xc magnetic field does not make a contribution
to spin dynamics, which is thus governed only by the noninteracting 
currents and the external magnetic field. In particular, for
perfectly localized noninteracting moments 
(such that ${\bf J}^{KS} \equiv 0$) and in the absence of an external
magnetic field (${\bf B}\equiv 0$), the ALDA yields
$d{\bf m}/dt \equiv 0$, so that there is no spin dynamics at all.
Within the linear-response approximation this problem of the ALDA has been
noticed previously in Ref.~\cite{jmmm3}.
The absence of spin dynamics for localized moments is another deficiency 
of the ALDA. This conclusion is consistent with the 
original phenomenological derivation of the noninteracting 
equation of motion \cite{lali}, which shows that the 
spin dynamics of localized moments is driven by {\it gradients} of the 
magnetization, and these are not contained in the ALDA.  

Typical LDA-based calculations of spin-dynamics
\cite{calc} avoid these problems by not proceeding 
entirely within the LDA (for example LDA is combined with constraining fields
which are not calculated selfconsistently, or it is used only to determine 
parameters in a model Hamiltonian which itself is not of the LDA form). 
Any first-principles calculation fully within the LDA, however, is bound 
to run into the above problems.

We thus now turn to a discussion of popular improvements upon the LDA. 
First, we note that any static functional that is invariant under the 
infinitesimal global spin rotation 
${\bf m}\r \to {\bf m}\r + \delta {\bf \varphi}\times{\bf m}\r$ satisfies the 
ZTT, when used as input for an adiabatic approximation to TD-SDFT.
An example is the exact-exchange (or OEP) method. 
Eq.~(\ref{ztt}) was derived only for exchange and correlation together, i.e.,
it is not guaranteed {\it a priori} that an exchange-only approximation
satisfies it. However, since the exact exchange term is rotationally
invariant, satisfaction of the ZTT is automatic. No similar result holds
for a general time-dependent, e.g. retarded, xc functional.

Next we consider gradient-dependent functionals. The static generalized
gradient approximations of Ref.~\cite{gga} depend only on $m_z$ and 
$\nabla m_z$. They also satisfy the ZTT identically, but at the price 
of making again ${\bf B}_{xc} \parallel {\bf m}$. A gradient-dependent
functional that depends on the full vector ${\bf m}$ and its derivatives
was sketched in Ref.~\cite{cdftlett} and another will be constructed below.
The general form of such functionals is
\be
E_{xc}=
\int d^3r\, e_{xc}(n,|{\bf m}|,\nabla n,\nabla m_x,\nabla m_y,\nabla m_z),
\label{ggaexample}
\ee
where the function $e_{xc}$ need not be specified explicitly for the present
purpose. 
Using the chain rule for functional 
derivatives we write the derivative of the generic functional 
(\ref{ggaexample}) as
\bea
-{\bf B}_{xc}\r = \frac{\delta E_{xc}}{\delta{\bf m}\r} =
\sum_k {\bf u}_k \frac{\delta}{\delta m_k\r}
\int d^3r'\, e_{xc}\rp
\nonumber \\
+ \sum_{ik} {\bf u}_k \int d^3x\,
\frac{\delta (\nabla m_i({\bf x}))}{\delta m_k\r}
\frac{\delta}{\delta (\nabla m_i({\bf x}))}
\int d^3r' e_{xc}\rp,
\eea
where the index $k$ labels components of the vector ${\bf m}$, and the unit
vector in direction of component $m_k$ is denoted ${\bf u}_k$.
The derivative in the first term on the right-hand side acts only on
$|{\bf m}|$, i.e., the terms $\nabla m_k$ are held fixed while differentiating.
By performing the derivatives of $\nabla m_i$ with respect to $m_k$ and
introducing the abbreviations
$f\r=
\delta/\delta|{\bf m}\r| \int d^3r'\, e_{xc}\r$
and
$ {\bf g}_k\r= \delta/\delta (\nabla m_k\r) \int d^3r' e_{xc}\r$
one obtains
\be
-{\bf B}_{xc}\r = {\bf u}_m f\r
- \sum_k {\bf u}_k \nabla \cdot {\bf g}_k\r,
\label{bxcfinal}
\ee
where ${\bf u}_m$ is the unit vector in direction of ${\bf m}$.
While the first term on the right-hand side is again parallel to ${\bf m}$,
the second, arising from the gradients in Eq. (\ref{ggaexample}), in general
is not. Independently of the detailed form of the unspecified kernel
$e_{xc}$, we thus find that gradient-dependent functionals of 
all three components of ${\bf m}$ produce a ${\bf B}_{xc}$ that is not 
necessarily parallel to ${\bf m}$.
Such functionals thus have the potential to solve the problems encountered by
the LDA: when used as input for an adiabatic approximation to TD-SDFT they can 
give rise to a nonzero ${\bf B}_{xc}^\perp\rt$ and hence, 
by virtue of Eqs. (\ref{tdsdftlifshitz}), (\ref{ltt}), and (\ref{bperp}), to 
nontrivial spin dynamics for localized moments, and to a nonzero 
${\bf J}_{xc}^L\rt$. 

However, the ZTT constrains the dependence of the functional on the
gradients of $m_x$, $m_y$ and $m_z$. In order to illustrate this with a simple 
example, relevant for the calculation of spin waves, 
we write ${\bf B}_{xc}\r=\int dr'\,\hat{K}[{\bf m},n]\rrp {\bf m}\rp$, 
where the kernel $\hat{K}\rrp$ is, for weakly inhomogeneous systems,
a short-range function of $|{\bf r}-{\bf r'}|$. We can then
expand ${\bf m}\rp$ to second order in $({\bf r}-{\bf r'})$,
\be
{\bf m}\rp  = {\bf m}\r +
({\bf r}-{\bf r'})\nabla\otimes {\bf m}\r +
({\bf r}-{\bf r'})^2\nabla^2{\bf m}\r,
\ee
and obtain
\be
{\bf B}_{xc}\r=\hat{K}_2{\bf m}\r + \hat{K}_4\nabla^2{\bf m}\r.
\label{bxcexpand}
\ee
Here $\hat{K}_2$ and $\hat{K}_4$ are proportional to the second and
fourth moments of the tensor $\hat{K}(|{\bf r}-{\bf r'}|)$, respectively.
{\it The ZTT is satisfied when these
moments, which in general are tensors, reduce to scalars.}
The first term in (\ref{bxcexpand}) then recovers the Stoner approximation
to the LDA, while the second is of the form of the terms appearing in the
phenomenological Lifshitz equation of motion for slowly varying moments 
(cf.~Eq.~(6.98) of Ref.~\cite{lali}).
Work to construct explicit expressions for these moments is in progress.

Up to this point we have considered conventional static SDFT only in so far 
as it provides
the input for adiabatic approximations to TD-SDFT. We now take a brief look at 
static SDFT in its own right. 
The equation of motion for ${\bf m}$ in static SDFT is simply
$d{\bf m}\r/dt = 0$.
Hence
\be
\hat{\nabla}\cdot {\bf J}^{KS}\r
=\frac{q}{2mc}{\bf m}\r  \times {\bf B}_s\r.
\ee
By comparing this with the corresponding many-body equation,
which differs from it only through the replacement of ${\bf B}_s$ by
the external field ${\bf B}$, and ${\bf J}^{KS}$ by ${\bf J}$, one
immediately finds
\be
\frac{q}{2mc}{\bf m}\r\times{\bf B}_{xc}\r =
\hat{\nabla}\cdot \left[ {\bf J}^{KS}\r -{\bf J}\r \right].
\label{statltt}
\ee
Since this is of the same structure as Eq.~(\ref{ltt}), our conclusions
(i) to (v) hold in static SDFT, too, however now applied to equilibrium
currents and magnetizations, and not to their dynamical counterparts.
Perhaps the most interesting result for static SDFT is the
existence of a component of ${\bf B}_{xc}$ perpendicular to ${\bf m}$,
since it implies that the prescription to take ${\bf B}_{xc}$ locally parallel 
to ${\bf m}\r$, which is employed in many recent calculations for noncollinear 
spin configurations \cite{noncoll} is consistent with 
Eq.~(\ref{statltt}) only to the extent that the right-hand side, i.e., the 
difference of the spin currents, can be neglected.

In summary, we have derived and analysed the equation of motion for the spin
degrees of freedom within TD-SDFT. 
Our results have consequences for, e.g., the calculation of spin currents 
in polymers and multilayers, the use of the ALDA in investigations of 
spin dynamics, popular methods for treating noncollinear magnetism, and the 
construction of better density functionals.

{\bf Acknowledgments}
KC thanks the Physics Departments of the University of Columbia, Missouri, USA,
and of the University of Bristol, UK, for generous hospitality, 
the FAPESP for financial support, and L.~N.~Oliveira and J.~C.~Egues for useful
discussions. 
This work resulted from a collaboration partially funded by the TMR network 
(Contract No. EMRX-CT96-0089).
GV acknowledges support from NSF Grant DMR-0074959.

\end{document}